\documentclass{IEEEtran}

\usepackage[T1]{fontenc}
\usepackage{graphicx}
\usepackage[colorlinks,urlcolor=blue,linkcolor=blue,citecolor=blue]{hyperref}
\expandafter\def\expandafter\UrlBreaks\expandafter{\UrlBreaks\do\/\do\*\do\-\do\~\do\'\do\"\do\-}
\usepackage{amsmath}

\makeatletter
\def\blfootnote#1{\begingroup\renewcommand\thefootnote{}\footnote{#1}\endgroup}
\makeatother

\begin{document}

\title{From Idea to Prototype in an Afternoon: Scaffolded, AI-Assisted Rapid VA Prototyping}

\author{Gennady~Andrienko~and~Natalia~Andrienko}

\markboth{IEEE Computer Graphics and Applications}{IEEE Computer Graphics and Applications}

\maketitle

\blfootnote{This is a revised manuscript (R1) submitted to IEEE Computer Graphics and Applications (Visualization Viewpoints department). Under review; posted as a preprint. \copyright~2026 IEEE. Personal use of this material is permitted. Permission from IEEE must be obtained for all other uses.}
\blfootnote{The authors are with Fraunhofer Institute IAIS, Sankt Augustin, 53757, Germany. E-mail: \{gennady|natalia\}.andrienko@iais.fraunhofer.de}

\begin{abstract}
Testing a new visual-analytics idea usually takes months: one needs to find a realistic data set, clean it, and implement an interactive prototype. We report a case where this effort was reduced to an afternoon, and use it to examine a loop we argue is central to AI-assisted prototyping: an assistant proposes a workflow and, later, an implementation, and an expert repeatedly assesses each proposal and gives corrective feedback. The idea under test: relax the Pareto frontier with a tolerance and group the surviving options into recurring types, ``constellations'' on a ``soft sky''. Using the Artifact-Transform Workflow Language (ATWL) as a scaffold, we obtained a sound workflow in minutes and a running prototype in hours, which let us observe the loop directly. At each stage the assistant proposed plausible methods quickly but could not judge which one was more appropriate. The first prototype was only average, and repeated expert feedback turned it into a state-of-the-art one. Controlled experiments confirm that scaffolds shape what the assistant proposes but do not remove the need for this judgment. We argue that the field needs a typology of human knowledge injection: what experts add, where in the loop, and why.
\end{abstract}

Having a new idea in visual analytics (VA) is easy; testing it is hard. A proper test may take months: a suitable data set needs to be found, cleaned, and understood, and an interactive prototype has to be implemented. Many potentially good ideas are never tried because this effort is too high. In this article, we use the term \emph{expert} for a VA researcher who wants to test and develop such ideas through prototype implementation.

\begin{figure}[t]
\centerline{\includegraphics[width=\columnwidth]{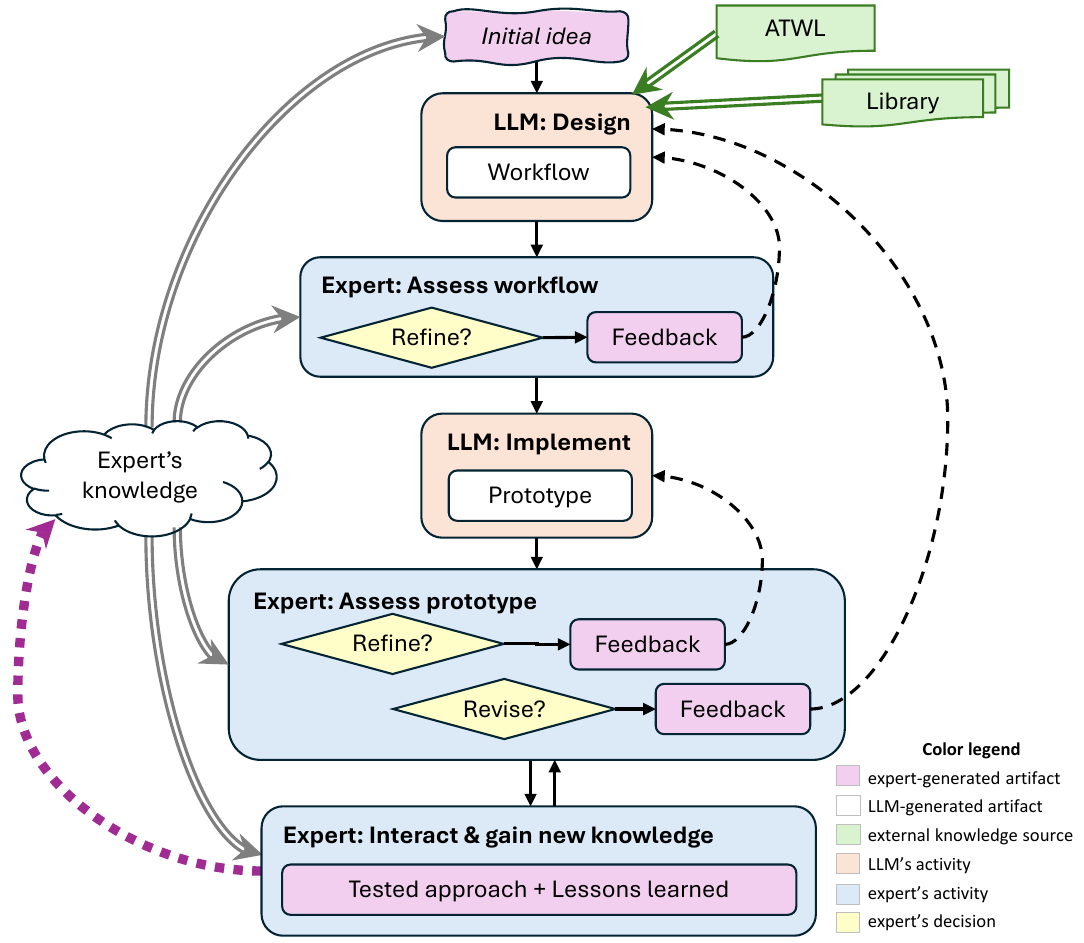}}
\caption{The AI-assisted prototyping loop.}
\label{fig1}
\end{figure}

We describe a case where this effort was reduced to an afternoon. For several years we had a concrete idea in mind: soften the Pareto frontier by introducing a tolerance, which produces a ``soft skyline'', and then group the surviving options into a small number of types metaphorically viewed as ``constellations on the soft sky'''. The idea was simple and we often discussed it, but we never implemented it. The required data preparation and prototyping did not justify the expected effort. This time we combined two tools that we had previously used separately: ATWL, an artifact-transform workflow language that we had developed,\cite{atwl} as a design scaffold, and a current large-language-model-based AI assistant (Claude Opus 4.8) as a ``build engine'' for producing code. Figure~\ref{fig1} summarizes this collaboration as a loop: the assistant proposes a workflow and, later, an implementation, and the expert repeatedly assesses each proposal and returns feedback that steers the next iteration. This loop, and where within it expert knowledge enters, is the subject of this article.

Multi-criteria decision analysis (MCDA) is a good test case. It is not present in ATWL's worked examples, and a design that combines a tolerant Pareto frontier with constellations is not part of the assistant's standard repertoire. The tools could not simply retrieve an existing solution. At each step, methods had to be chosen rather than copied, which made this case a realistic test of the combined approach.

\begin{figure*}[t]
\setlength{\fboxsep}{6pt}%
\noindent\fbox{\begin{minipage}{\dimexpr\textwidth-2\fboxsep-2\fboxrule\relax}
\footnotesize
\textbf{SIDEBAR 1. Multi-Criteria Decision Analysis}\\[3pt]
Multi-criteria decision analysis (MCDA) helps a decision maker choose among options rated on several conflicting criteria. Most visual tools descend from two ideas. The first is the \emph{value function}: criteria are mapped to a common desirability scale and combined into a single rank. A weighted sum is the simplest rule; richer aggregations - ordered weighted averages, ideal-point and outranking methods - give finer control, especially in spatial decision problems [S1]. Interactive tools such as LineUp [S2], Podium [S3], and WeightLifter [S4] let analysts explore how weights change the ranking. Such ranking is \emph{compensatory}: a high score on one criterion can offset a low score on another, making sensitivity exploration through coordinated views essential. A recent human-centered strand makes preference elicitation itself interactive, letting non-experts score and weight mixed attributes and refine the resulting ranking [S5].\\[3pt]
The second idea is \emph{Pareto dominance}: the skyline keeps every option that no other option beats on all criteria at once [S6], assuming no weights. The skyline is sensitive to measurement error and grows quickly with the number of criteria, so in high dimensions it filters little. A large literature relaxes it - $\varepsilon$-dominance [S7] and $k$-dominant skylines [S8] - to control the count, though usually by \emph{discarding} options.\\[3pt]
SoftSky sits between the two families: it relaxes dominance by a per-criterion tolerance that \emph{admits} near-optimal options rather than removing them, then groups the survivors into a few interpretable, named types.\\[6pt]
\textbf{References}\\[2pt]
{[S1]} N.~Andrienko and G.~Andrienko, ``Informed spatial decisions through coordinated views,'' \textit{Information Visualization}, vol.~2, no.~4, pp.~270--285, 2003.\newline
{[S2]} S.~Gratzl, A.~Lex, N.~Gehlenborg, H.~Pfister, and M.~Streit, ``LineUp: Visual analysis of multi-attribute rankings,'' \textit{IEEE Trans.\ Vis.\ Comput.\ Graphics}, vol.~19, no.~12, pp.~2277--2286, 2013.\newline
{[S3]} E.~Wall, S.~Das, R.~Chawla, B.~Kalidindi, E.\,T.~Brown, and A.~Endert, ``Podium: Ranking data using mixed-initiative visual analytics,'' \textit{IEEE Trans.\ Vis.\ Comput.\ Graphics}, vol.~24, no.~1, pp.~288--297, 2018.\newline
{[S4]} S.~Pajer, M.~Streit, T.~Torsney-Weir, F.~Spechtenhauser, T.~M\"oller, and M.~Waldner, ``WeightLifter: Visual weight space exploration for multi-criteria decision making,'' \textit{IEEE Trans.\ Vis.\ Comput.\ Graphics}, vol.~23, no.~1, pp.~611--620, 2017.\newline
{[S5]} C.-M.~Barth, J.~Schmid, I.~Al-Hazwani, M.~Sachdeva, L.~Cibulski, and J.~Bernard, ``How applicable are attribute-based approaches for human-centered ranking creation?,'' \textit{Comput.\ Graphics}, vol.~114, 2023.\newline
{[S6]} S.~B\"orzsönyi, D.~Kossmann, and K.~Stocker, ``The skyline operator,'' in \textit{Proc. Int.\ Conf.\ Data Eng.}, 2001, pp.~421--430.\newline
{[S7]} M.~Laumanns, L.~Thiele, K.~Deb, and E.~Zitzler, ``Combining convergence and diversity in evolutionary multiobjective optimization,'' \textit{Evol.\ Comput.}, vol.~10, no.~3, pp.~263--282, 2002.\newline
{[S8]} X.~Lin, Y.~Yuan, Q.~Zhang, and Y.~Zhang, ``Selecting stars: The $k$ most representative skyline operator,'' in \textit{Proc.\ IEEE 23rd Int.\ Conf.\ Data Eng.\ (ICDE)}, 2007, pp.~86--95.
\end{minipage}}
\end{figure*}

This loop yields three observations, which structure the rest of this article. First, the combination is fast: we obtained a sound workflow in minutes and a working prototype in hours. Second, the scaffold matters: without ATWL, the same class of assistant produced a clearly weaker, naive design. A controlled experiment across two tasks turns this observation into a practical recommendation on scaffold use. Third, and central to this article, the scaffold is not sufficient: the first implementation achieved only average quality, and we had to inject expert knowledge at several points to reach state-of-the-art quality. ATWL is not designed to contain this kind of knowledge, and the assistant did not reliably supply it.

This difference between an average prototype and a good one came from method-level decisions made by a human expert: which technique to use, how to encode results, and in which order to apply the steps. This knowledge exists mainly in experts' heads and is scattered over many publications. We argue it should be represented in a form that is both human-editable and machine-accessible, and that the community needs a typology of such human knowledge injection. We base this argument on a detailed account of one development process, relating it to VA as model building\cite{modelbuilding} and visualization for human-LLM partnership.\cite{collab}

\section{A LONG-HELD IDEA AND A TEST}

\textbf{The Idea}. Standard tools for multi-criteria decisions have well-known limitations (Sidebar~1). A weighted score is compensatory: strong criteria can hide weak ones. The Pareto skyline is brittle: differences within measurement error can change dominance relations, and in high dimensions almost nothing is dominated, so the skyline contains many options and stops filtering effectively.
Our idea lies between these approaches. We soften dominance by a tolerance, so that the frontier becomes a band rather than a hard line - a soft sky. We then interpret the band not as a flat list but as a small number of groups of similarly good options - constellations on the frontier. The constellations are found automatically under human supervision and then interpreted and named.

We had sketched this idea for years but never implemented it. A fair test requires a data set that is large and imperfect enough to stress the method and an interactive prototype that allows analysts to explore the results. For us, the expected effort for data preparation and prototyping was always too high.

To perform a fair test, we chose a public data set of German apartment-rental offers \url{https://www.kaggle.com/datasets/corrieaar/apartment-rental-offers-in-germany}, with hundreds of thousands of listings and realistic problems: out-of-range values, missing fields, duplicate re-postings, and no given coordinates. We focused on Leipzig, the large city whose offers contain the most complete street addresses and can be geocoded most reliably.
Cleaning and geocoding the Leipzig subset followed standard procedures: a ten-step pipeline and address matching against an open database. The result was about 13,000 offers, most at house-number resolution.

\section{THE METHOD: CONSTELLATIONS ON A SOFT SKY}
The method implemented by the prototype has value in its own right, so we describe it precisely enough to be reproduced. Each option $o$ is rated on $m$ criteria, and a utility function maps each raw value to a desirability $u_k(o)\in[0,1]$, so that a larger value is better on every criterion.

\textbf{The Soft Sky.}
An option remains on the frontier unless some rival is decisively better. In the classical skyline, a rival $q$ dominates $o$ if $q$ is at least as good as $o$ on every criterion and strictly better on at least one. The non-dominated options form the skyline.

We make ``decisively better'' require a margin. For each criterion $k$ we fix a tolerance $\varepsilon_k\ge 0$, which is the smallest difference on criterion $k$ that we treat as real rather than noise. Then $q$ \emph{tolerantly dominates} $o$ only if it beats $o$ by at least that margin on \emph{every} criterion:
\begin{equation}
u_k(q)\;\ge\;u_k(o)+\varepsilon_k \qquad (k=1,\dots,m).
\end{equation}
An option \emph{survives} (lies on the soft sky) when no rival passes this test against it. The tolerances thus act as a bar that a rival must clear in order to win. Raising $\varepsilon_k$ raises this bar, so fewer rivals clear it, fewer options are dominated, and the soft sky \emph{grows}. For $\varepsilon_k=0$ the test reduces to the ordinary skyline; as the $\varepsilon_k$ grow, the frontier widens.

This direction is deliberate and opposite to approximate $\varepsilon$-dominance in multiobjective optimization, where a tolerance is used to \emph{thin} the frontier. Here it is used to \emph{widen} it, so that near-optimal options are kept rather than discarded.
For each option we compute a signed robustness margin
\begin{equation}
\rho(o)\;=\;\min_{q\neq o}\;\max_{k}\;\frac{u_k(o)-u_k(q)}{\varepsilon_k},
\end{equation}
with the minimum taken over all other options (with $\varepsilon_k>0$). Intuitively, $\rho(o)$ measures how well $o$ withstands its most threatening rival. We have $\rho(o)>-1$ exactly when $o$ lies on the soft sky; $\rho(o)\ge 0$ for classically non-dominated options; and $-1<\rho(o)<0$ for options admitted only because of the tolerance. Larger $\rho(o)$ means that the option lies deeper inside the frontier. The margin is a single continuous optimality measure, computed in $O(n^2 m)$ time for $n$ options.

\textbf{Constellations.}
The soft sky can still contain many options, so we group them into recurring types metaphorically called ``constellations'''. For each option we discretize its utilities into a few ordered levels per criterion (for example, low, medium, and high). This turns the option into a set of \emph{(criterion, level)} tokens.

We then treat options as documents and tokens as terms and apply nonnegative matrix factorization (NMF) $\mathbf{V}\approx\mathbf{W}\mathbf{H}$ with $r$ factors. This recovers $r$ recurring combinations of criteria levels. Each option is assigned to the factor for which it has the strongest weight, and each constellation is named by the tokens that contribute most to its factor (for example, \emph{cheap, central, small}). Membership is defined directly by the factorization, so a constellation and its name cannot diverge. Figure~\ref{fig3} shows a fragment of the prototype interface.

\section{FAST: A WORKFLOW IN MINUTES, A PROTOTYPE IN HOURS}
We used ATWL as a scaffold (Sidebar~2).\footnote{Supplementary material, including the SoftSky workflow expressed in ATWL, interactive prototype and demo walkthrough (online): \url{https://geoanalytics.net/softsky}}  Together with the LLM assistant, we designed the workflow plan before writing any code: focus the spatial extent of the options, choose criteria, define each criterion's utility, set a tolerance, compute the soft sky, group its members into constellations, and support decision making. The artifacts and their flow, three refinement loops, and a feedback path that carries evidence from the frontier back to the settings were all fixed in minutes.

\begin{figure}[t]
\centerline{\includegraphics[width=18.5pc]{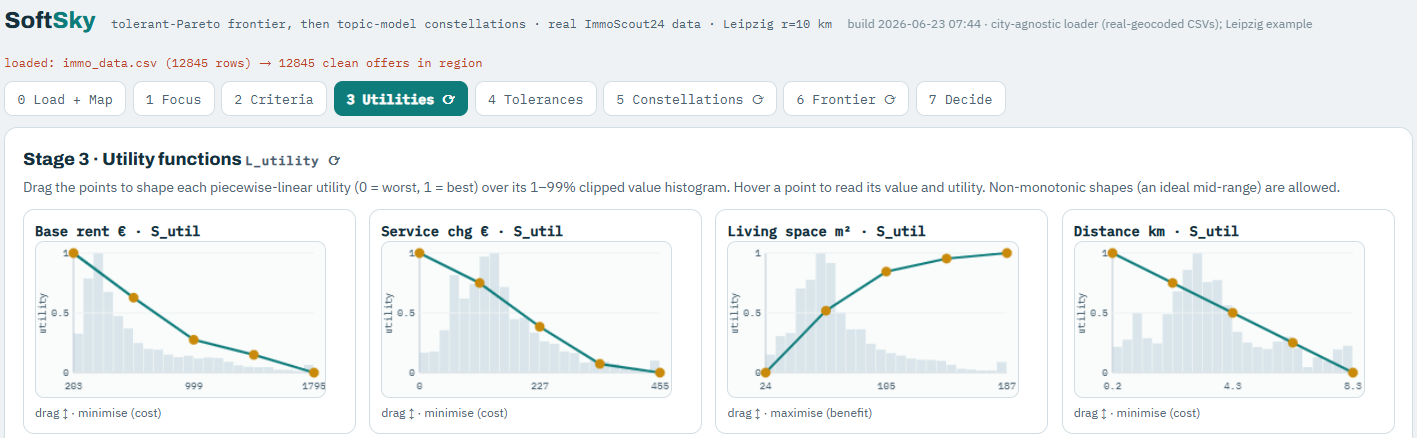}}
\caption{Prototype at \url{geoanalytics.net/softsky} (fragment).}
\label{fig3}
\end{figure}

The scaffold helped in two ways. The language gave the workflow its structure, and the library of examples suggested typical architectural patterns: staged narrowing, dual decomposition, and other ideas we would otherwise have introduced from scratch. ATWL kept every step typed and made refinement loops explicit. Much of the speed came directly from re-using these patterns.

\begin{figure}[t]
\setlength{\fboxsep}{6pt}%
\noindent\fbox{\begin{minipage}{\dimexpr\columnwidth-2\fboxsep-2\fboxrule\relax}
\footnotesize
\textbf{SIDEBAR 2. The Artifact--Transform Workflow Language}\\[3pt]
ATWL describes a visual-analytics workflow as a directed graph of typed \emph{artifacts} joined by \emph{transforms}.\cite{atwl} Artifacts carry the state of an analysis: \emph{entities} and their \emph{features} (the data), \emph{models} fitted to them, \emph{patterns} read from data or models, externalized \emph{knowledge}, and \emph{specifications} that parameterize later steps. Each transform names the artifact types it consumes and produces and the analytic \emph{intent} it serves, so a workflow records what was computed and why. The language was created to describe and compare published workflows. It operationalizes the model-building view of visual analytics\cite{modelbuilding} and comes with a library of worked cases.\\[3pt]
Used forward, before any code is written, ATWL becomes a design scaffold. It forces every product of a step to be a typed artifact rather than an informal sketch. It makes refinement loops and feedback paths first-class objects, so iteration becomes part of the design. It also locates each step at a level of the nested model of visualization design,\cite{munzner} separating \emph{what} a step does from \emph{how} it is implemented. That separation is both the strength and the limit of ATWL. It fixes the \emph{shape} of a workflow: its artifacts, their types, and their flow. It says nothing about which algorithm, encoding, or parameter should be used in a given step. These choices are left to the builder.
\end{minipage}}
\end{figure}
\begin{figure}[t]
\setlength{\fboxsep}{6pt}%
\noindent\fbox{\begin{minipage}{\dimexpr\columnwidth-2\fboxsep-2\fboxrule\relax}
\footnotesize
\textbf{SIDEBAR 3. AI-Assisted VA Prototyping}\\[3pt]
A growing body of work uses large language models to shorten the path from design to a working visual-analytics system. Recent surveys map LLM roles across the analytic pipeline [S1]; agent-based frameworks such as LightVA plan tasks and assemble linked views without requiring deep programming skills [S2]; and toolkits driven by natural language turn requests into chart specifications with justified choices [S3, S4]. These systems can assemble code and views quickly. Our case study demonstrates that the harder method-level choices - which technique to use, which encoding to choose, and how to order the steps - still require expert knowledge injection.\\[6pt]
\textbf{References}\\[2pt]
{[S1]} M.~Hutchinson et al., ``LLM-assisted visual analytics: Opportunities and challenges,'' in \textit{Proc.\ CGVC}, 2024.\newline
{[S2]} Y.~Zhao et al., ``LightVA: Lightweight visual analytics with LLM agent-based task planning and execution,'' \textit{IEEE Trans.\ Vis.\ Comput.\ Graphics}, vol.~31, no.~9, pp.~6162--6177, 2025.\newline
{[S3]} A.~Narechania, A.~Srinivasan, and J.~Stasko, ``NL4DV: A toolkit for generating analytic specifications for data visualization from natural language queries,'' \textit{IEEE Trans.\ Vis.\ Comput.\ Graphics}, vol.~27, no.~2, pp.~369--379, 2021.\newline
{[S4]} L.~Wang et al., ``LLM4Vis: Explainable visualization recommendation using ChatGPT,'' in \textit{Proc.\ EMNLP}, 2023.
\end{minipage}}
\end{figure}

The implementation followed quickly. The assistant generated a self-contained web page with a map, controls for criteria and utilities, tolerance, constellations, and ranking over a few hours of interaction. Because the scaffold had fixed the structure, each exchange revised a method or encoding rather than the workflow skeleton. By the end of the session we had a running prototype on real Leipzig data (Figure~\ref{fig3}): we could vary the tolerance, inspect constellations, and rank options within a chosen constellation. The slow part of testing an idea from a blank page to a first functional prototype on real data took less than one working day, not weeks.

Using an LLM as the build engine is itself an active research direction (Sidebar~3). Several toolkits and agent frameworks now assemble code and linked views from a prompt. This is the Design and Implement side of the loop in Figure~\ref{fig1}: fast, but everything it proposes should be treated as still just a proposal. The rest of this article follows the Assess side of the loop, where expert judgment enters. As the next two sections show, that judgment is what turns an average prototype into a good one.

\section{THE SCAFFOLD MATTERS}
To isolate the scaffold's contribution, we ran a simple counterfactual. We gave another LLM the same task as initially (design an interactive workflow for choosing among many options rated on conflicting criteria, keeping near-optimal options and summarizing the good ones into a few kinds) but without any ATWL material. The comparison with the first workflow is qualitative. Prompts, models, and a structured side-by-side comparison are in the supplement online.

The assistant generated a noticeably weaker design, suggesting to filter the options, project them into two dimensions, let the user draw groups on the projection, then pick criteria, construct a frontier, and rank options. The crucial weakness is the grouping step. A group that is drawn on a two-dimensional projection is bound to one particular view and one random seed, relies on a geometry that is known to distort neighborhood relationships,\cite{tsne} and has no precise definition that can be re-applied or audited. The naive design also tended to compute a frontier inside each group. This let the grouping step influence what counts as non-dominated, a wrong dependency.

Designing in ATWL avoided these traps. The language is not ``clever'', but it forces every product of a step to be a typed artifact. A hand-drawn region is, in ATWL terms, a pattern abstracted from a visualization: it cannot be edited, re-applied, or fed forward as a specification. The artifact that the workflow needs is a specification, and stating this explicitly pushes the design toward groups that are defined in the data space, computed once, and named automatically. The scaffold thus turned a vague ``group the options'' into a typed step with clear inputs and outputs. This is our second observation: the scaffold was what prevented a fast workflow from becoming a naive one.

\section{THE SCAFFOLD IS NOT ENOUGH: EXPERT KNOWLEDGE INJECTION}
The first prototype worked, but it was clearly mediocre. Turning it into something we would describe as state-of-the-art required a second, longer phase, and almost all of the additional work focused on the methods and encodings under the workflow, the layer that ATWL types but does not fill. For each open slot the assistant itself proposed some method, but most of these first proposals were later replaced. Each replacement was a case of knowledge injection: at a specific workflow step, the expert overrode the assistant's default with method knowledge that the assistant did not reliably possess. Five such injections had a major impact on quality. Each follows the same pattern.

\noindent\textbf{1. Direction of the tolerance.} The assistant implemented the textbook approximate $\varepsilon$-dominance, which prunes the frontier towards a few representative points. This is the opposite of what our idea requires. We inverted the rule to the growing form of Equation~(1) and added the signed margin of Equation~(2).

\noindent\textbf{2. How to group.} The assistant factorized the continuous matrix of criterion values, producing groups that were hard to interpret as weighted combinations of raw numbers. We changed to topic modeling over discrete criterion levels, so that a constellation appears as a recurring combination of named levels.

\noindent\textbf{3. How to define a group.} To define groups more precisely, the assistant suggested decision-tree rules. However, these rules only approximated the clusters. We dropped the rules and let the model's own factor assignments decide which options belong to each constellation. The only manual step was to check and possibly edit the name for each constellation.

\noindent\textbf{4. Role of the embedding.} The assistant proposed a two-dimensional embedding to define groups by outlining them, the same default behavior as in the counterfactual without ATWL. We demoted the embedding to a single inspection view and defined constellations in the original criterion space, where they are reproducible and do not depend on the embedding's distortions.\cite{tsne}

\noindent\textbf{5. Where the frontier is computed.} The assistant computed a frontier inside each group. We inverted the order: we first computed one global frontier and then grouped its surviving options. In this way the grouping describes the frontier instead of redefining it.

Across the five injections, the expert judged by a small, recognizable set of \emph{criteria}: whether a method matched the stated design intent, whether its result was interpretable, whether it was reproducible independent of an arbitrary choice such as a projection's seed, and whether the workflow's steps depended on each other in the right order. Four injections were made only at ``Assess prototype'' in Figure~\ref{fig1}, once the prototype existed. The ATWL workflow specification was correct at the abstract level. Checking whether a proposed method inside a slot was the right one was hard to do by reading the specification alone, but easy once the prototype made the method's behavior visible. Only the choice of how to define a group (injection 3) was caught earlier, at ``Assess workflow''. Two injections, the tolerance direction and the grouping method, changed only the implementation. The other three changed the workflow itself: which artifact fed into which step, or in what order the steps ran.

Each of these injections corresponds to knowledge that an experienced visual-analytics researcher typically has, e.g. that an embedding distorts global structure, and that an approximate rule for a cluster can be a handicap. In this process, the assistant was a strong proposer and a weak judge. It recalled and implemented methods quickly and could even prove small properties of them when asked, but it could not reliably distinguish a good method from a merely plausible one. The expert could, and that judgment, not the assistant's fluency, moved the prototype from average to good. Because MCDA is not a worked example in ATWL and the tolerant-Pareto-with-constellations design has no standard implementation to copy, each of these decisions had to be made rather than retrieved.

Fast iteration also changed how we thought about the method, not just how quickly we obtained a prototype. Because the prototype responded in seconds, it became an instrument for reasoning about the method itself. We could formulate a conjecture, have it implemented, and see whether it holds within minutes. For example, we asked how the global soft sky is related to two-dimensional views of pairs of criteria. Experiments showed that lying on the soft sky of any single pair is a \emph{sufficient} condition for global survival but not a \emph{necessary} one: an option that is non-dominated in one scatterplot is safe, yet an option can survive globally while appearing dominated in every pairwise view.

Iteration also helped core design ideas to crystallize. The decision to let the frontier \emph{grow} as the tolerance increases (the inversion at the core of the method) emerged from observing how the surviving set reacted to the slider, not from a derivation on paper. Used in this way, the fast loop contributed to the development of the method. It did not merely produce a demonstration. Some refinements were not knowledge that we already had and simply inserted. They were \textit{new knowledge that we formed in the loop}.

Experimenting also revealed limits of the method. A single multiplicative $\varepsilon_k$ is suitable for continuous criteria but not for integer ones such as the number of rooms. For such criteria a fractional tolerance is meaningless and a fixed threshold (for example, ``within one room'') is a more appropriate relaxation.

\section{VALIDATING THE SCAFFOLD}
So far the claim that the scaffold matters rests on one counterfactual. To test it systematically, we ran controlled experiments across two unrelated tasks: the apartment MCDA workflow above and a second workflow for COVID-19 pandemic behavior analysis. These experiments target one question: how do the two ATWL scaffolds, the language definition and the library of examples, shape what the assistant proposes? They report scaffold effects relevant to the knowledge-injection argument of this article. ATWL as a language, independent of LLM use, is described in full elsewhere.\cite{atwl} The results below confirm what the abstract states: scaffolds shape proposals, but they do not remove the need for expert assessment.

At the core is a $2{\times}2$ factorial design crossing the two scaffolds ($\pm$\,ATWL language definition $\times$ $\pm$\,ATWL library of worked examples), yielding four single-pass conditions per task, run in parallel with identical prompts. We also ran multi-pass strategies to test whether sequential prompting avoids the limitations found in the single-pass runs. We qualitatively rated each output on three dimensions: coverage of analytical content, formal correctness, and architectural sophistication. We used no blind protocol or second rater and ran all conditions on a single model, so the scores are structured impressions rather than a controlled measurement. Full conditions, prompts, outputs, and scoring are online.

\smallskip\noindent\textbf{Single-pass results.}
Three findings held across both tasks. (1) Without any scaffold, the model was most creative but least formal: only the unscaffolded output contained usage scenarios, method-comparison tables, and mathematical formulations. (2) The two scaffolds serve different roles. The language definition acts at the token level: it improves formal correctness but narrows conceptual exploration. The library acts at the phase level: it encourages more sophisticated analytical architectures without enforcing syntactic compliance. (3) Providing both together did not produce the best result. Output was competent but ordinary, as if the model's attention had split between two attractors. A plausible explanation is a zero-sum token budget: a formal template in context fills output space with easy-to-produce declarations, displacing harder-to-generate content such as justifications and scenarios.

\smallskip\noindent\textbf{Multi-pass prompting.}
To test whether sequential prompting avoids this displacement, we ran a three-pass strategy (an unconstrained design pass, then a library-inspired refinement, then a formalization pass introducing the language) and a two-pass strategy (an unconstrained design pass, then a refinement pass with both scaffolds together) on both tasks. Both removed the displacement effect: creative concepts from the first pass persisted into formalization. The two-pass strategy was the more token-economical of the two, reaching the same coverage in fewer interactions. But token cost is not the only cost that matters. Each additional pass is also a chance for the expert to give feedback and steer the next step, so for genuine design work one may decide to accept the extra passes.

\smallskip\noindent\textbf{Implications.}
These experiments refine our earlier observations. The scaffold matters, but not as a single tool: the language disciplines a workflow's form, and the library enriches its architecture. Neither replaces expert knowledge injection. In both studies, the library condition produced the most sophisticated structure yet still missed the method-level judgments described above, such as the tolerance direction and the reduced role of embeddings, which only the expert supplied.

The displacement effect suggests a practical rule: introduce scaffolds after an initial creative pass, not simultaneously with the first design prompt. The scaffold works best when there is already something to steer, and its output is still a proposal for the expert to assess.

\section{TOWARD A TYPOLOGY OF HUMAN KNOWLEDGE INJECTION}
Our argument continues the knowledge-assisted visualization tradition, which separates tacit from explicit knowledge and models the conversions between them\cite{KnowledgeConversion}. This tradition treats explicit knowledge as a first-class container on the machine side, one that can be externalized and then exploited to steer analysis, visualization, and guidance\cite{ExplicitKnowledge}. The five injections done in our case study are externalizations in exactly this sense, with one shift of target: they address the \emph{building} of a VA system rather than the analysis of a data set.

Figure~\ref{fig1} locates these injections within the loop introduced earlier. ATWL scaffolds the Design stage: it fixes what the steps are and how they connect. The assistant fills Design and Implement quickly, proposing candidate methods and encodings on demand. Neither the scaffold nor the assistant covers the decisions that had the largest impact on quality: which method is appropriate for a given slot under the conditions at hand. As the five injections showed, the expert made most of these decisions at ``Assess prototype'', once a prototype existed to judge, and occasionally already at ``Assess workflow''. It was these injections, not the scaffolds, that made the final result strong.

The process to study is therefore \textit{knowledge injection} itself: what an expert adds, where in the loop this happens, and in what form the knowledge appears. The five injections described earlier already suggest several types: a fitness judgment that selects one technique over another; an encoding choice that makes a result legible; an ordering constraint between steps; a decision to use a tool only for inspection rather than for definition; and a reversal of a default direction. A useful typology would name these kinds of contributions, locate them within the loop, and record their conditions of use and possible failure modes.

This task fits well with ongoing work in visualization for machine learning (Vis4ML) and human-centered AI (HCAI), which already aim to structure how human knowledge enters machine-assisted analysis.\cite{sacha,modelbuilding} The new element is that this knowledge should now also be accessible to an AI assistant during system construction, not only during analysis.

The term ``injection'' is somewhat misleading: it can make an impression that knowledge is fixed beforehand and only applied. In our case, knowledge did not simply flow from a finished store in the expert's head into the prototype. It also evolved during the loop of design, implementation, and inspection. Some contributions were pre-existing expertise, for example, that embeddings distort global structure or that topic models over discrete levels stay readable. Other contributions were \emph{generated} in the loop, in the way the tolerance-direction and pairwise-frontier findings described above were: not recalled, but discovered by watching the prototype respond. The fast prototype was a \emph{medium for thinking}, not just a target to configure. Each update raised a question, and its answer became new knowledge that influenced the next update.

A typology of human contribution should therefore distinguish between knowledge that is \emph{applied} and knowledge that is \emph{formed}. The first type covers importing an established method into a suitable slot. The second type covers refining, combining, or inventing methods in response to what the prototype reveals. This second type is more difficult to capture in advance, but it is also where the scientific value lies.

A structured representation of injected knowledge would need to differ from what we already have. A curated prompt library stores useful phrasings but not the conditions under which a method is appropriate. A decision tree or flowchart fixes the routing in advance, whereas method choices are often contested and revised as the field develops. General ontologies such as VIS4ML\cite{sacha} catalogue entities and stages of VA-assisted ML describe \emph{what exists}, but not the fit conditions and failure modes that determine \emph{what to use}.

What appears to be missing is something between a pattern language and an executable knowledge base. Its entries would be human-readable, like design guidelines, but also typed and queryable. A workflow language could then refer to them, and an assistant could be guided by their conditions of use. Capturing knowledge injection has three requirements.

\textbf{1. Representation.} Each method needs a typed entry: what it does, which artifact types it consumes and produces, the conditions under which it is a good choice, and the expected failure modes.

\textbf{2. Editability.} The entries must be human-readable and easy to correct, because they encode evaluative judgments that are still being developed. A static hand-curated catalogue is too slow to update, and an opaque model is not correctable. The envisioned target form should be close to the explicit knowledge structures that VA has long advocated.\cite{modelbuilding}

\textbf{3. Accessibility.} The same entries must be reachable by an assistant at the moment it proposes a method, so that choices are guided by recorded fit conditions rather than training-data frequencies. This is the role that recent work on human-LLM partnership envisages for externalized, machine-readable knowledge.\cite{collab} Here, it applies to method choices made while building a system, rather than to the provenance of analyses.

In practice, entries could be short, and many are already known: (1) use topic modeling over discrete levels when groups must be human-readable; (2) avoid continuous factorization when the factors must be explained; (3) use a neighbor embedding to inspect but not to define groups. A workflow language supplies the slots. A typology of knowledge injection, together with such entries, would help both experts and assistants decide what to put into those slots.

\section{LIMITATIONS}
This article describes a single case, not a validation study. While the resulting method, constellations on a soft sky, is interesting in its own right, the main contribution was to reveal a phenomenon: the need for expert steering and knowledge injection in AI-assisted prototyping. Our article aims to define a research agenda around that phenomenon, not to claim that ATWL and Claude Opus 4.8 are the best scaffold and assistant for this task. The following conditions limit the applicability of the lessons we can draw.

The afternoon timescale assumed an expert who already knew which methods were likely to work. A newcomer to MCDA would have to invest time to learn the methods that the assistant implemented incorrectly. The expert in this case was a VA methods researcher, not a domain expert in rental markets or MCDA end-use. Validating the constellation metaphor with real decision-makers is a separate question this case does not answer. The case also assumed a problem that can be expressed in ATWL's artifact types and a method whose building blocks already exist in software libraries, not a genuinely new algorithm.

Finally, we have described expert knowledge injection mainly as a correction of the assistant. Experts have biases too. An injected choice can be wrong, or right only in a limited context. A shared body of injected knowledge will need the same scrutiny, versioning, and discussion as any other shared resource.

These boundaries do not overturn the main finding: expert judgment, not the scaffold or the model, determined the final quality. This is what a typology of knowledge injection would need to test and extend.

\section{CONCLUSION}
We set out to test an idea that we had carried for years, and a workflow language plus an AI assistant allowed us to do so in an afternoon rather than over several weeks. The case makes a clear distinction between three contributions: (1) The scaffold did real work: ATWL made the workflow sound and well typed, and its library of cases suggested a suitable architecture. (2) The AI assistant made the implementation fast. (3) Expert judgment made the result good.

Our experiment refines the role of the scaffold. The language and the library play different roles and work best when they are applied sequentially after a first creative pass. Fewer passes save tokens, but more passes give the expert more opportunities to inject knowledge and improve the design. The main structure of the argument remains unchanged. The first two contributions are now relatively cheap. The third is not. The knowledge that it depends on is not yet available in a form that the first two can use.

We therefore expect that the next substantial gain in this style of prototyping will come not from a new workflow language or a larger model, but from a typology of human knowledge injection (what experts add, where in the loop they add it, and why) captured so that experts can edit it and assistants can use it\cite{InjectKnowledge}.

\textbf{ACKNOWLEDGMENTS.}
The authors are grateful to Piotr Jankowski, J\"urgen Bernard, and Michael Sedlmair for extensive discussions and collaboration on MCDA- and ATWL-related topics. This work was done within the Lamarr Institute for Machine Learning and Artificial Intelligence.

\def\refname{REFERENCES}

\begin{IEEEbiography}{Gennady Andrienko}
is a lead scientist at Fraunhofer Institute IAIS, Sankt Augustin, Germany, a professor at City St George's, University of London, U.K., and a co-PI at the Lamarr Institute for Machine Learning and Artificial Intelligence, Germany.
\end{IEEEbiography}

\begin{IEEEbiography}{Natalia Andrienko}
is a lead scientist at Fraunhofer Institute IAIS, Sankt Augustin, Germany, a professor at City St George's, University of London, U.K., and a co-PI at the Lamarr Institute for Machine Learning and Artificial Intelligence, Germany.
\end{IEEEbiography}


\begin{thebibliography}{9}

\bibitem{atwl}
N. Andrienko, G. Andrienko, J. Bernard, and M. Sedlmair, ``ATWL: a formal language for representing, comparing, and reusing Visual Analytics workflows,'' arXiv:2605.25489, 2026.

\bibitem{modelbuilding}
N. Andrienko et al., ``Viewing visual analytics as model building,'' {\it Comput. Graphics Forum}, vol. 37, no. 6, pp. 275--299, 2018.

\bibitem{collab}
M. Elshehaly et al., ``Designing for collaboration: Visualization to enable human-LLM analytical partnership,'' {\it IEEE Comput. Graphics Appl.}, vol. 45, no. 5, pp. 107--116, 2025.

\bibitem{tsne}
M. Wattenberg, F. Vi\'egas, and I. Johnson, ``How to use t-SNE effectively,'' {\it Distill}, 2016.

\bibitem{KnowledgeConversion}
X. Wang et al., ``Defining and applying knowledge conversion processes to a Visual Analytics system,'' {\it Comput. Graphics}, vol. 33, no. 5, pp. 616--623, 2009.

\bibitem{ExplicitKnowledge}
P. Federico et al., ``The role of explicit knowledge: a conceptual model of knowledge-assisted Visual Analytics,'' {\it IEEE VAST}, pp. 92--103, 2017.

\bibitem{munzner}
T. Munzner, ``A nested model for visualization design and validation,'' {\it IEEE Trans. Vis. Comput. Graphics}, vol. 15, no. 6, pp. 921--928, 2009.

\bibitem{sacha}
D. Sacha, M. Kraus, D. A. Keim, and M. Chen, ``VIS4ML: An ontology for visual analytics assisted machine learning,'' {\it IEEE Trans. Vis. Comput. Graphics}, vol. 25, no. 1, pp. 385--395, 2019.

\bibitem{InjectKnowledge}
Y. Xing et al., ``Understanding how humans inject knowledge into machine learning workflows through Visual Analytics''. arXiv:2607.00969, 2026.

\end{thebibliography}
\end{document}